\documentclass[preprint,aps,prb,superscriptaddress,showpacs]{revtex4}

\usepackage{graphics}
\usepackage{epsfig}
\usepackage{color}
\usepackage{amsmath}
\begin{document}

\title{One particle spectral function and analytic continuation for many-body implementation
in the EMTO method}

\author{A. \"Ostlin}
\email[Corresponding author. E-mail: ]{ostli@kth.se}
\affiliation{Department of Materials Science and Engineering, Applied Materials Physics, KTH Royal Institute of Technology, Stockholm SE-100 44, Sweden} %
\author{L. Chioncel}
\affiliation{Augsburg Center for Innovative Technologies, University of Augsburg, D-86135 Augsburg, Germany}%
\affiliation{Theoretical Physics III, Center for Electronic Correlations and Magnetism, Institute of Physics, University of Augsburg,
D-86135 Augsburg, Germany}
\author{L. Vitos}
\affiliation{Department of Materials Science and Engineering, Applied Materials Physics, KTH Royal Institute of Technology, Stockholm SE-100 44, Sweden} %
\affiliation{Department of Physics and Astronomy, Division of Materials Theory, Uppsala University, Box 516, SE-751210, Uppsala, Sweden} %
\affiliation{Research Institute for Solid State Physics and Optics, Wigner Research Center for Physics, Budapest H-1525, P.O. Box 49, Hungary} %

\date{11 October 2012}


\begin{abstract}
We investigate one of the most common analytic continuation techniques in condensed matter physics, namely the Pad\'{e} approximant. Aspects concerning its implementation in the exact muffin-tin orbitals (EMTO) method are scrutinized with special regard towards making
it stable and free of artificial defects. The electronic structure calculations are performed for solid hydrogen, and the performance of the analytical continuation is assessed by monitoring the density of states constructed directly and via the Pad\'{e} approximation. 
We discuss the difference between the \textbf{k}-integrated and \textbf{k}-resolved analytical continuations, as well as describing the use of random numbers and pole residues to analyze the approximant. It is found that the analytic properties of the approximant can be controlled by appropriate modifications, making it a robust and reliable tool for electronic structure calculations. At the end, we propose a route to perform analytical continuation for the EMTO + dynamical mean field theory (DMFT) method.
\end{abstract}
\maketitle

\section{Motivation}
Dynamical mean field theory combined with density functional theory formulated within
 the local density approximation (LDA+DMFT) is a rapidly developing area of the modern computational solid state physics
 of strongly correlated materials. A number
of different implementations have been proposed and applied to calculate 
spectral functions and thermodynamic properties of Mott insulators, ferromagnetic 
3$d$, 4$f$  metals, actinides or other systems~\cite{ko.sa.06,ka.ir.08}. Most of the 
LDA+DMFT calculations have been performed with partial self-consistency applied 
only to the many-body local self-energy, while the LDA potential (Hamiltonian)  
was kept fixed. In such schemes the effect of electronic correlations upon the 
charge is neglected. Recently, full self-consistent LDA+DMFT schemes became also available \cite{sa.ko.04,ch.vi.03,mi.ch.05}. 
The first implementation by Savrasov et al. \cite{sa.ko.04} uses the full-potential linear muffin-tin orbitals (LMTO) method basis set,
the many-body self-energy is added directly to the Hamiltonian, and the diagonalization
of a non-Hermitian problem is performed. Within the multiple scattering approach,
the first implementation was realized within the EMTO basis set \cite{ch.vi.03},
followed by the KKR implementation \cite{mi.ch.05}. While in the EMTO+DMFT approach the 
self-energy is added to the scattering path operator, from which the charge density 
is obtained by a complex contour integration, in the latter KKR+DMFT approach the
self-energy is added directly to the radial Dirac solver from where the regular
and irregular solutions are computed directly. In both these EMTO and KKR Green function 
schemes, the LDA solver is formulated within the complex plane. Therefore the combination
with the many-body problem requires the ``transfer'' from the complex plane to the 
imaginary axis or the real axis where the many-body problem is solved. In both
these implementations, the Pad\'{e} analytical continuation is used to promote the Green's 
function from the complex contour into the many-body solver. 

In the present paper, we discuss Pad\'{e} analytical continuation schemes, 
in order to clarify their accuracy in the EMTO calculations. We propose a new approach for an accurate scheme which is verified using the EMTO one-particle Green's function. Numerical tests are performed 
on solid hydrogen, which is a simple $s$-band system.  

The paper is organized as follows: in the next section we review the major implementation steps of the EMTO+DMFT method and establish the one-particle Green's function that is subject to the analytical 
continuation. In Sec. \ref{pade}, we present details of the Pad\'{e} analytical 
continuation and in Sec. \ref{revpade} we discuss its major implementation issues. In Sec. \ref{proposenumtest} we propose procedures to test its accuracy and consider the example of hydrogen as a prototype $s$-band system to be investigated. Conclusions are given in Sec. \ref{conclusion}.

\section{Overview of the EMTO method}
\label{emto+dmft}

Within the multiple-scattering formalism, the one-electron Green's function $G^{\sigma}({\bf r},
{\bf r'},z)$ is defined for an arbitrary complex energy $z$ as

\begin{equation} \label{oeGf}
\left[z+\nabla^2_{\bf r}-v{^\sigma}_{eff}({\bf r})\right] G^{\sigma}({\bf r},
{\bf r'},z)\;=\;\delta({\bf r-r'}),
\end{equation}
where $\sigma$ stands for the spin, and ${\bf r}$ and ${\bf r'}$ are the position coordinates.
 For most of the applications, e.g. standard KKR or LMTO methods, the LSDA effective
potential from Eq.\ (\ref{oeGf}) is approximated by spherical muffin-tin (MT) wells
centered at lattice sites $R$. Within a particular basis set, the one-electron Green's
function is expressed in terms of the so-called scattering path operator,
$g^{\sigma,{\rm LSDA}}_{RL,R'L'}(z)$, as well as the regular,
$Z^{\sigma}_{RL}(z,{\bf r}_R)$, and irregular, $J^{\sigma}_{RL}(z,{\bf r}_R) $,
solutions to the single site scattering problem for the cell potential at lattice
site $R$, {\it viz.}

\begin{eqnarray}\label{g1}
G^{\sigma,{\rm LSDA}}({\bf r}_R+{\bf R},{\bf r}_{R'}+{\bf R'},z)&=&
\sum_{L,L'}Z^{\sigma}_{RL}(z,{\bf r}_R)g^{\sigma,{\rm LSDA}}_{RL,R'L'}(z)
Z^{\sigma}_{R'L'}(z,{\bf r}_{R'})\;-\nonumber\\
&-&\delta_{RR'}\sum_LJ^{\sigma}_{RL}(z,{\bf r}_R)Z^{\sigma}_{RL}(z,{\bf r}_R),
\end{eqnarray}
where $L\equiv(l,m)$ with $l<l_{max}$ (usually $l_{max}=3$) and ${\bf r}_R\equiv
{\bf r}-{\bf R}$ denotes a point around site $R$. The real space representation for
the scattering path operator for the muffin-tin potential is given by

\begin{equation}\label{gsmall}
g^{\sigma,{\rm LSDA}}_{RL,R'L'}(z)\;=\;
[\delta_{RR'}\delta_{LL'}{t^{\sigma}}^{-1}_{RL}(z)-B_{RL,R'L'}(z)]^{-1},
\end{equation}
where $t^{\sigma}_{RL}(z)$ stands for the single scattering $t$-matrix and
$B_{RL,R'L'}(z)$ are the elements of the so-called structure constant matrix.

Conventional MT based KKR or LMTO methods have limited accuracy as a result of 
the shape approximation to the potential and charge density. The former
method uses non-overlapping spherical muffin-tin potentials and constant potential
in the interstitial, while the latter method approximates the system with overlapping
atomic spheres neglecting completely the interstitial and the overlap between
individual spheres. Recent progress in the field of {\it muffin-tin orbitals}
theory \cite{an.sa.00,an.je.94,vito.01,vi.sk.00,vito.07} shows that the best possible representation of the full
potential in terms of spherical wells may be obtained by using large overlapping
muffin-tin wells \cite{an.ar.98,zw.an.09,vito.07} with properly treated overlaps. Within this so-called {\it exact
muffin-tin orbitals} method \cite{an.sa.00,an.je.94,vito.01,vi.sk.00,vito.07}, the scattering path operator $g^{\sigma,{\rm LSDA}}_{RL,R'L'}(z)$ is calculated as the inverse of the kink matrix defined by

\begin{equation}\label{g2}
K^{\sigma}_{RL,R'L'}(z) \equiv
\delta_{RR'}\delta_{LL'}D^{\sigma}_{RL}(z) - {S}_{RL,R'L'}(z),
\end{equation}
where $D^{\sigma}_{RL}(z)$ denotes the EMTO logarithmic derivative function
\cite{vi.sk.00,vito.01}, and ${S}_{RL,R'L'}(z)$ is the slope matrix \cite{an.sa.00}.
Since the energy derivative of the kink matrix, ${\dot K}^{\sigma}_{RL,R'L'}(z)$,
gives the overlap matrix for the EMTO basis set \cite{an.sa.00}, the matrix
elements of the properly normalized LSDA Green's function become \cite{vi.sk.00,vito.01}

\begin{equation}\label{green}
G^{\sigma,{\rm LSDA}}_{RL,R'L'}(z)\;=\;
\sum_{R''L''}g^{\sigma,{\rm LSDA}}_{RL,R''L''}(z){\dot K}^{\sigma}_{R''L'',R'L'}(z) -
\delta_{RR'}\delta_{LL'} I^{\sigma}_{RL}(z),
\end{equation}
where $I^{\sigma}_{RL}(z)$ accounts for the unphysical poles of
${\dot K}^{\sigma}_{RL,R'L'}(z)$. In the case
of translation invariance, Eqs.\ (\ref{g2}) and (\ref{green}), are transformed to the
reciprocal space, so that the lattice index $R$ runs over the atoms in the primitive
cell only, and the slope matrix, the kink matrix, and the path operator depend
on the Bloch wave vector ${\bf k}$.


The total number of states at the
Fermi level $E_F$ is obtained as

\begin{equation}\label{nos}
N(E_F) = \frac{1}{2\pi i}\sum_{RL,R'L'}\oint
\int_{BZ}G^{\sigma,{\rm LSDA}}_{RL,R'L'}({\bf k},z)\;d{\bf k}\;dz,
\end{equation}
where the energy integral is carried out
on a complex contour that cuts the real axis below the bottom of the valence
band and at $E_F$. The \textbf{k}-integral is performed
over the first Brillouin zone, which numerically is turned into a sum over a discrete {\bf k}-mesh in the irreducible Brillouin zone (IBZ). 

\begin{figure}[h!]
	\includegraphics[scale=1]{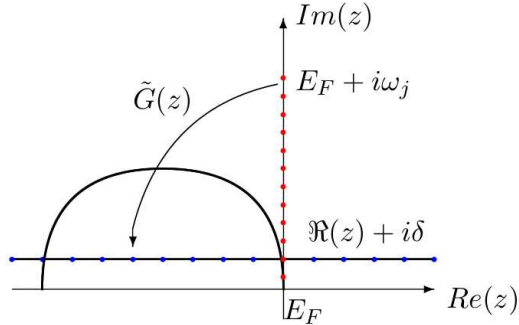}
	\caption{Schematic picture of the complex energy contours used in the EMTO+DMFT method. The charge density and number of states are calculated for points $z$ on a semi-circle enclosing the valence band. For finite temperature calculations (e.g. LDA+DMFT) the equidistant Matsubara points $i\omega_j$ (in red) along the imaginary axis relative to the Fermi level $E_F$ are used. The density of states is calculated for points $z$ (blue) on a horizontal contour defined as $\Re(z)+i\delta$ ($\delta$ being a small real number).}
	\label{fig1}
\end{figure}

In the EMTO+DMFT method \cite{ch.vi.03}, the many-body effects are added to the DFT-level Green's function through a local self-energy ${\tilde \Sigma}_{RL,RL'}^{\sigma}(z)$ via the Dyson equation
\begin{equation}  \label{G0}
\left[G_{RL,R'L'}^{\sigma}({\bf k},z)\right]^{-1} =
\left[G_{RL,R'L'}^{\sigma,{\rm LSDA}}({\bf k},z)\right]^{-1} -
\delta_{RR'}{\tilde \Sigma}_{RL,RL'}^{\sigma}(z).
\end{equation}
where $G_{RL,R'L'}^{\sigma}({\bf k},z)$ is now the LSDA+DMFT Green's function matrix, computed on the complex contour. The ${\bf k}-$integrated LSDA+DMFT Green's function, $G_{RL,R'L'}^{\sigma }(z)=\int_{BZ}G_{RL,R'L'}^{\sigma}({\bf k},z)d{\bf k}$, is analytically continued $G(z) \xrightarrow{\mathrm{Pad\acute{e}}} G(i\omega)$ to the Matsubara frequencies $\omega _{j}=(2j+1)\pi T$, where $j=0,\pm1,...$, and $T$ is the temperature. From this latter Green's function the {\it bath} Green's function is computed according to
\begin{equation} \label{Gtilde}
\left[{\cal G}_{RL,R'L'}^{\sigma }(i\omega)\right]^{-1} =
\left[G_{RL,R'L'}^{\sigma }(i\omega)\right]^{-1}+
\delta_{RR'}{\tilde \Sigma}_{RL,RL'}^{\sigma}(i\omega).
\end{equation}
The analytical continuation is performed by the Pad\'{e} method \cite{ch.vi.03}. The many-body problem
is solved on the Matsubara axis, and the resulting self-energy is then analytically continued to the semi-circular contour $\tilde{\Sigma}(i\omega) \xrightarrow{\mathrm{Pad\acute{e}}} \tilde{\Sigma}(z)$. In Figure \ref{fig1} we illustrate the contours used in the EMTO+DMFT calculations. Accordingly, the self-consistency procedure requires two analytic continuation steps, that has to be controlled numerically.

In the present work we proceed as follows: the self-consistent LSDA Green's function is computed on the complex contour in the usual procedure as shown above. From the self-consistent potential we compute explicitly Eq. (\ref{g2}) and Eq. (\ref{green}) on the Matsubara frequencies. The Green's function is then analytically continued from the imaginary axis to a horizontal contour, described in Figure \ref{fig1}, and detailed analysis of accuracy of such a scheme is performed.

\section{Overview of Pad\'{e} approximants}
\label{pade}
An analytic function $f(z)$ can be uniquely specified by values contained in a compact set. Construction of a sequence of
 rational functions $\tilde{f}(z)$ having the same values as $f(z)$ on a given set of points is known as
the rational interpolation problem.

The Pad\'{e} approximant method is a possible way to solve the rational interpolation 
problem, and has proved very useful in providing quantitative information
about the solution of many interesting problems in physics and chemistry.

One of the most used techniques to construct the Pad\'{e} approximant is the Thiele reciprocal difference method
which was pioneered in the field of condensed matter physics by Vidberg and Serene \cite{vi.se.77} who used it as a 
means to analytically continue spectral functions from the imaginary axis 
to the real axis. The method of Thiele takes values $f(z_1), f(z_2), ...,f(z_{N})$ from $N$ (complex) points 
$z_1, z_2, ..., z_{N}$ as input and returns an analytically continued function value $\tilde{f}(z')$ at a point $z'$. 
This is performed with $O(N^2)$ complexity. If $N$ is an odd number $\tilde{f}$ will be a rational polynomial 
function where both the numerator and denominator will be of order $(N-1)/2$. If $N$ is even the denominator 
polynomial will be of order $N/2$ while the numerator will be of order $N/2-1$. Hence, if the goal is to
approximate a spectral function with $M$ number of poles, then $N=2M$ input points should be used. In general 
$M$ is not known \emph{a priori}. To resolve this issue Vidberg and Serene \cite{vi.se.77} observed that a good approximant can be expected by increasing $N$ until the number of approximant poles in the lower half of the complex plane converges to a fixed value.
 
Later on, the Pad\'{e} approximant method was scrutinized by Beach et al. \cite{be.go.00} who introduced the matrix 
formulation, which allows to compute the rational polynomial coefficients from which the approximant
is constructed. As these coefficients are obtained by a matrix inversion the issue of accuracy is addressed   
using a multi-precision arithmetic in the calculation of the rational polynomial. 
In that study 
the approximant
\begin{equation}\label{poly}
\tilde{f}(z) = \frac{a_1 + a_2z + \cdots + a_Mz^{M-1}}{b_1 + b_2z + \cdots + b_Mz^{M-1} + z^M}
\end{equation}
was calculated explicitly by solving a linear system of $N$ ($N$ even) equations to get the complex polynomial coefficients $\{a_i,b_i\}$. In the following section we will present several issues concerning the use of the Pad\'{e} method in electronic band structure calculations, in particular for the EMTO implementation. 

\section{Revised Pad\'{e} approximation}\label{revpade}
\subsection{{\bf k}-integrated vs. {\bf k}-resolved analytical continuation}\label{kvsk}

\begin{figure}[h!]
\includegraphics[scale=1]{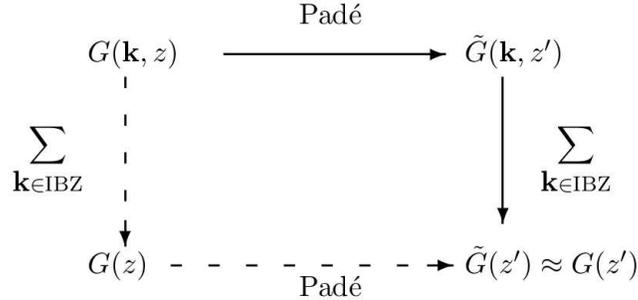}
\caption{Schematic diagram of the analytic continuation before (straight path) or after (dashed path) Bloch summation. $G(\textbf{k},z)$ in the top left corner is given for a set of points. The goal is to obtain an analytic expression $\tilde{G}(z')$ (lower right corner) that can be used to approximate the true function $G(z')$. Following the straight path, the Pad\'{e} approximant needs to be constructed for each $\textbf{k} \in$ IBZ and the Bloch sum is performed at the desired point. Along the dashed path, only one single Pad\'{e} approximant is needed.}
\label{kvsloc}
\end{figure}

In the case of the \textbf{k}-resolved Green's function in Eq. (\ref{G0}), excluding ${\tilde \Sigma}_{RL,RL'}^{\sigma}(z)$, we have some 
\emph{a priori} knowledge of the pole structure. This is because each orbital contributes with 
at least one pole (hybridization between bands will in general give 
rise to more than a single pole). On the other hand, the pole structure of the local Green's function 
in Eq. (\ref{Gtilde}) is not known explicitly. Since the ideal Pad\'{e} approximant of a function $f$ by construction is a rational polynomial, it has a clearly defined pole structure. Hence performing the analytic continuation on the \textbf{k}-resolved Green's function should be more suitable. 
The two different procedures are shown schematically in Figure \ref{kvsloc}, and we focus in this Article on the spectral properties (density of states) generated in these procedures, leaving out the self-energy. 
In the next subsection we discuss pole-zero pairs of the approximant in the complex plane.

\subsection{Pole-zero pairs of the approximant}\label{pzpairs}
An important issue is the nearly canceling pole-zero pairs in the Pad\'{e} approximant.
These pairs arise when the order of the denominator in Eq. (\ref{poly}) is larger than the actual number of poles present in the function to be analytically continued. In the spirit of the matrix formulation of Beach et al. \cite{be.go.00},
in such cases the linear equations used to solve the rational interpolation problem becomes overcomplete. If this happens, the 
Pad\'{e} algorithm tries to cancel the spurious poles by placing a zero of the numerator at the location of 
the pole. The pole-zero cancellation will not be exact due to the lack of arbitrary precision. 
These pole-zero pairs are defects in the approximant, since they are not part of the true function. Hence, 
these pairs should be found by performing a root-search on both the denominator and nominator polynomial and 
once found be removed from the approximant. Assuming that we have found a set of $k$ pole-zero pairs 
$\{(p_1,q_1),(p_2,q_2), ...,(p_k,q_k)\}$, where the $p_i$'s are zeros and the $q_i$'s are poles of the 
approximant, the pairs can be filtered out by the following operation
\begin{equation}\label{remfroi}
\tilde{f}(z) \rightarrow \tilde{f}(z) \prod^k_i \frac{z-q_i}{z-p_i}.
\end{equation}
In general it is difficult to make any clear statement on where exactly in the complex plane these pole-zero pairs
will appear. There are theorems about the overall distribution \cite{baker.75}, however for the present problem they are too general to be of any practical use.

\subsubsection{Geometric search of true poles}

A pertinent question is how to decide which poles are physical and which are spurious. The most 
obvious way would be to make a geometric search in the complex plane and investigate whether a chosen pole 
has a zero in its neighborhood. One of the simplest ways would be to let the neighborhood be a circle of some predefined radius. This straightforward method has some pitfalls. It is for example not obvious how to specify the neighbourhood in the most optimal way since the pole-zero pairs usually have different separations, as will be seen in Sec. \ref{proposenumtest}. If the neighborhood is defined too large then more than one zero can be present 
inside leading to ambiguity or the accidental removal of a physical zero. If on the other hand the neighborhood is chosen too small, a spurious pole in a pole-zero pair with a large separation could mistakenly be considered as a true pole of the approximant. Hence, in practice, this method is difficult to implement inside any self-consistent loop, where an algorithm would be necessary to automate the filtering of defects.

\subsubsection{Selection of true poles by multiplication with random numbers}
Another approach to select the true poles is to add a complex random number to the original input data, which was done by Sokolovski et al. for S-matrix elements \cite{so.ak.11}. 
The reasoning behind this method is that the physical poles should be insensitive to this (numerical) perturbation while the 
spurious pole-zero pairs should redistribute themselves in the complex plane.
Adding a random number to the input Green's function would effectively lead to adding an extra zero to the nominator, affecting the asymptotic properties of the Green's function. In this study, we chose to multiply the input Green's function with real positive random numbers, in order to conserve the asymptotic properties of the function. Notice that multiplicative positive noise only leads to a rescaling of the pole weights, but keeps the asymptotic properties intact.
Accordingly, the input Green's function is multiplied with real positive random numbers of the form 

\begin{equation}
1-\eta x_i, \;\; i=1, ..., N,
\end{equation}
where $\eta$ is a real scaling factor and the $x_i$'s are real random numbers between $-0.5$ and $0.5$.

\subsubsection{The approximant as a sum of poles}\label{wall}
One can learn more about the Pad\'{e} approximant by writing it as a product of its poles and zeros (second to lhs Eq. (\ref{defwi})), and as a sum of its poles and residues (rhs Eq. (\ref{defwi})) viz.
\begin{equation}\label{defwi}
\tilde{f}(z) = C \frac{\prod\limits^{M-1}_{i=1}(z-p_i)}{\prod\limits^{M}_{i=1}(z-q_i)} =
\sum\limits^{M}_{i=1}\frac{w_i}{z-q_i}
\end{equation}
where $w_i$ is the residue (or \emph{weight}) of the $i$'th pole. $C$ is a multiplicative complex constant that needs to be added back when reassembling the approximant as a product of its poles and zeros. This is because the root-finding procedure is invariant up to a scaling factor for the polynomial coefficients. The weights  $w_i$ can be found by some algebra. Multiply Eq. (\ref{defwi}) with $(z-q_j)$
\begin{equation}
C(z-q_j)\frac{\prod\limits^{M-1}_{i=1}(z-p_i)}{\prod\limits^{M}_{i=1}(z-q_i)} = 
(z-q_j)\sum\limits^{M}_{i=1}\frac{w_i}{z-q_i}.
\end{equation}
After some rearrangement we arrive at
\begin{equation}
w_j = C\frac{\prod\limits^{M-1}_{i=1}(z-p_i)}{\prod\limits^{M}_{i \neq j}(z-q_i)} - 
(z-q_j)\sum\limits^{M}_{i \neq j}\frac{w_i}{z-q_i}.
\end{equation}
If we now set $z=q_j$ we finally get that
\begin{equation}\label{wiqi}
w_j = C\frac{\prod\limits^{M-1}_{i=1}(q_j-p_i)}{\prod\limits^{M}_{i \neq j}(q_j-q_i)}.
\end{equation}
Notice that all the poles have to be non-degenerate to avoid dividing by zero. Extension to degenerate poles will be considered in a future study. It is now possible to make a connection with two theorems (Theorem 43.1 in \cite{wall.48}, and Ref. \cite{wa.we.44}) stating the relationship between a sum of the form in the right-hand side of Eq. (\ref{defwi}) and its analytic properties. These theorems imply that if all the $w_i$'s are positive and all the $q_i$'s are real, the approximant has positive-definite, integrable spectral weight and is analytic except along the real axis. In the spirit of Beach et al. \cite{be.go.00}, this is what is needed for the spectral function to be physical. Using the above information,
one can single out the physical poles from the defective ones, and write the approximant as
\begin{equation}\label{wipi}
\tilde{f}(z) = \sum_i{}^{'} \frac{w_i}{z-q_i},
\end{equation}
where the prime indicates that the sum should be taken over physical poles only. Which poles that are true physical poles or defects should be decided on the ground of the above theorems. For example, if a pole has negative residue it can be considered unphysical and be removed. The Pad\'{e} method will in general assign a
non-zero imaginary part to the residues due to lack of arbitrary precision, even to the physical poles as is shown in Sec. \ref{proposenumtest}. Hence the exclusion of poles from the approximant due to non-zero imaginary parts has limited use. 

\section{Proposed new scheme and examples}\label{proposenumtest} 
After consideration of the above issues, we propose and test the following scheme to perform the Pad\'{e} analytic continuation of the Green's functions in the EMTO+DMFT method:
\begin{itemize}
\item Calculate the \textbf{k}-resolved LSDA Green's function (the integrand in Eq. (\ref{nos})) and construct a Pad\'{e} approximant for each \textbf{k}-point. Perform the Bloch sum only after the analytic continuation has been performed.
\item Search for physical poles by multiplication of a random positive numbers to the original input data and see how the poles and zeros redistribute themselves in the complex plane. In parallel, investigate the properties of all approximant pole weights.
\end{itemize}
The idea behind the proposed scheme is to perform analytic continuation on  \textbf{k}-resolved Green's functions. 
After the construction of the Pad\'{e} approximant, the root-finding procedure is performed on the denominator and numerator polynomials using a Weierstrass-Durand-Kerner-Dochev method \cite{so.ak.11,pe.ca.95}. After the roots are
found the pole-zero pairs are explicitly removed from the approximant. 

In the following we perform several numerical tests to validate the proposed method.
To compare approximated and exactly calculated spectral functions we consider the density of states as a function of energy $DOS(E)=-\frac{1}{\pi}Im[G(E+i0^+)]$. Using the EMTO Green's function explicitly calculated at the first $N$ Matsubara frequencies, we use the Pad\'{e} method to
analytically continue this function to points situated on a horizontal contour defined as $z \equiv \Re(z)+i\delta$, where $\delta$ is a real constant (see Figure \ref{fig1}). The Pad\'{e} approximant is constructed by the newly proposed method to obtain the rational polynomial form in Eq. (\ref{poly}). To analyze the density of states we introduce an error, as a function of the energy $E$, viz.
\begin{equation}\label{relerr}
Error(E) = | DOS(E)-DOS_{app}(E) |
\end{equation}
as a measure of the deviation between the directly calculated density of states $DOS(E)$ and the approximant $DOS_{app}(E)$ obtained from the Pad\'{e} form.

\subsection{Solid hydrogen - a single $s$-band}

We consider hydrogen in the face-centered cubic structure. The Green's function written in the EMTO basis will carry an $s$-character, avoiding any hybridization with higher $\ell$-states. In the actual calculation, the Wigner-Seitz radius is set to 2.50 Bohr and the Bloch sum is performed using 10,569 \textbf{k}-points in the irreducible Brillouin zone. The EMTO equation is solved for 16 complex energy points along a semi-circular integration contour with a depth of 1.0 Ry relative to the Fermi level, enclosing the valence states. The slope matrix in Eq. (\ref{g2}) is calculated using the two-center expansion \cite{ki.vi.07}, and the local density approximation \cite{pe.wa.92} is used for the exchange-correlation functional. After self-consistency is reached the DOS is calculated as outlined above using a contour with $\delta=0.02$ Ry.

\subsubsection{\textbf{k}-integrated Pad\'{e} approximation}

\begin{figure}[h!]
	\includegraphics[scale=0.5,clip=true]{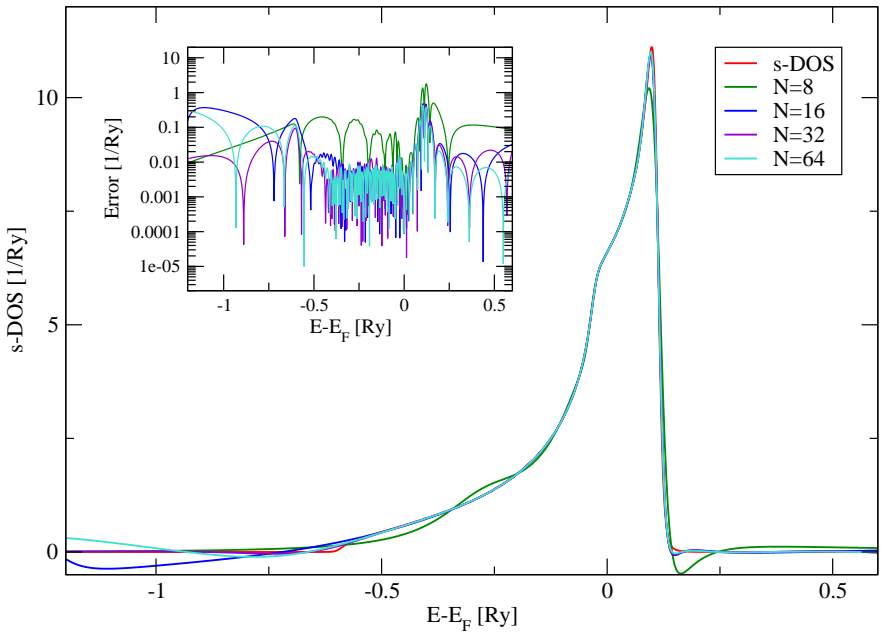}
	\caption{Density of states for solid hydrogen compared with approximants. The approximants were
	constructed using the $\ell m$-resolved local Green's function, using the $N$ first Matsubara points
	corresponding to a temperature of $500$ K. Inset: $Error$ as defined in Eq. (\ref{relerr}).}
	\label{hydlmdos}
\end{figure}

First we investigate the Pad\'{e} approximant for the \textbf{k}-integrated Green's function, i.e. the procedure illustrated by the dashed path in Figure \ref{kvsloc}.
The results obtained by analytically continuing the local Green's function $G_{RL,R'L'}(i\omega_j)$, ($L=L'=\{\ell=0,m=0\}$, $R=R'$ single site), using the first $N=8,16,32$ and $64$ Matsubara points evaluated for a temperature $T=500$ K are shown in Figure \ref{hydlmdos}. It can be noted that the Pad\'{e} approximant can capture the smooth behavior of the DOS around the Fermi level, but it has difficulties in correctly reproducing the function at the band edges. Even though the approximant finds the correct location of the top of the band at roughly 0.1 Ry above the Fermi level, it either under- or overshoots the exact DOS while, at the same time, it introduces negative spectral weight.
In order to improve the approximation we investigate the \textbf{k}-resolved Pad\'{e} approximant.

\subsubsection{Pad\'{e} approximation for a single \textbf{k}-point}

\begin{figure}[h!]
    \includegraphics[scale=0.5,clip=true]{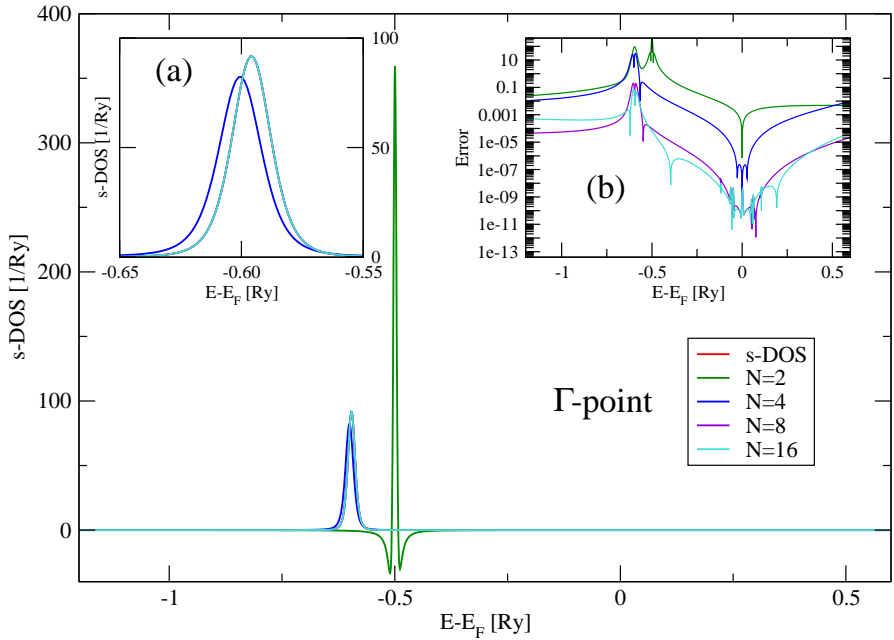}
	\caption{Density of states for the $\Gamma$-point compared with approximants of different order.
	 Inset (a): Same as main Figure, but zoomed in around $-0.65$ to $-0.55$ Ry. Inset (b): $Error$ as defined in Eq. (\ref{relerr}).}
	\label{gammor}
\end{figure}

Next we investigate the Pad\'{e} approximant for a single \textbf{k}-point. For illustration purposes, here we select the $\Gamma$-point and test the Pad\'{e} continuation for the Green's function calculated for $\mathbf{k}=(0,0,0)$. For the present test case (fcc H), this spectral function has a physical pole at about -0.6 Ry below the Fermi level (corresponding to the bottom of the $s$ band, DOS of $\Gamma$-point shown in Figure \ref{gammor}). For the N=2 approximant the single approximant pole is placed at about $-0.50+i0.01$ Ry below the Fermi level, giving a bad fit. As the order of the approximant increases, the approximant produces a better fit. The need of a higher approximant is motivated by the nature of the multiple-scattering Green's function used in the EMTO method, that encodes different scattering states. Within the EMTO basis set that is restricted in this calculation to the $s$-states the projected Green's function generates only a single pole.

\begin{figure}[h!]
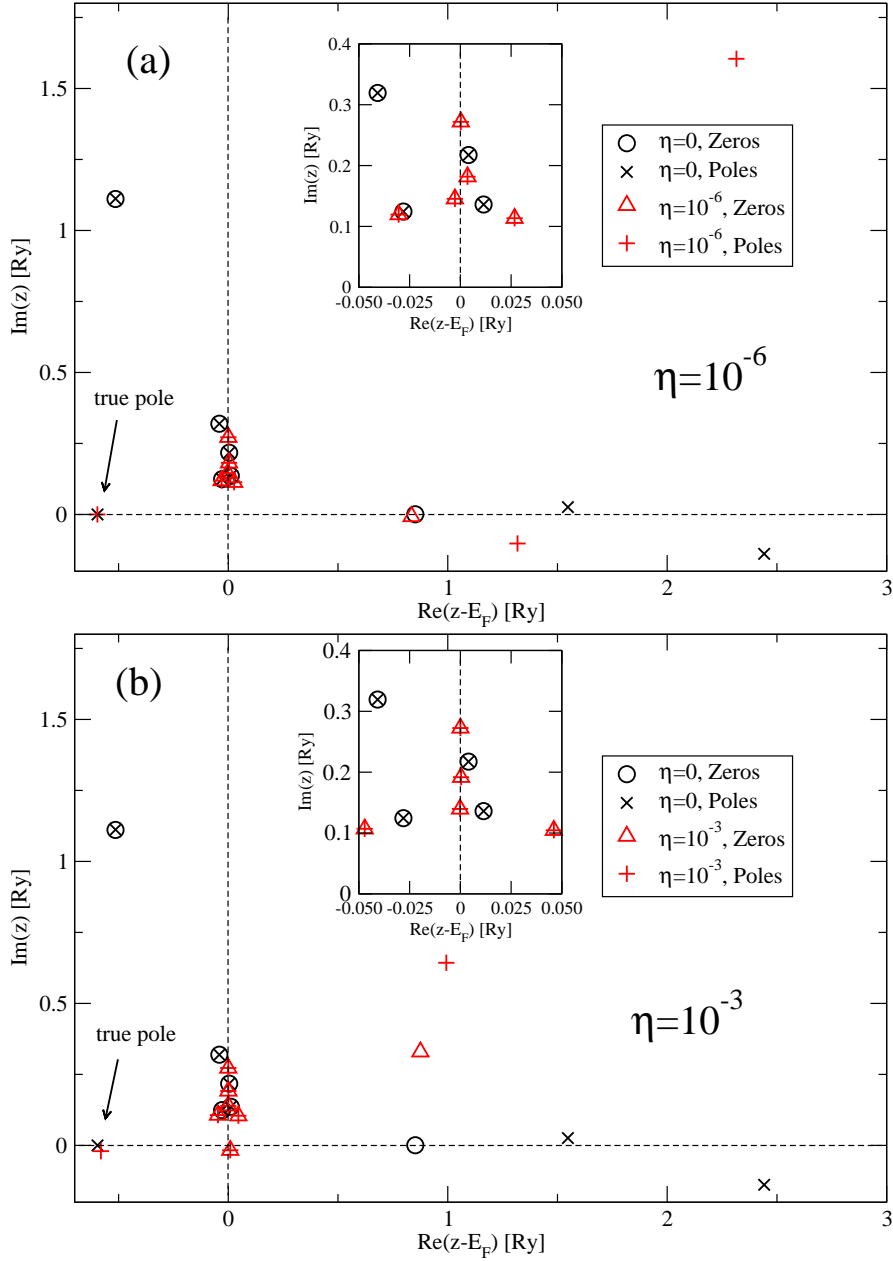

    \includegraphics[scale=0.5,clip=true]{fig5.eps}
    \includegraphics[scale=0.5,clip=true]{fig6.eps}
	\caption{Top (a): Pole-zero distribution of the $N=16$ approximant for the $\Gamma$-point, with and without randomness. The true
	pole stemming from the original input data is situated at $-0.5959-i3.5719\cdot10^{-6}$ Ry. Adding randomness 
    with $\eta=10^{-6}$ gives the pole position at $-0.5963+i3.4499\cdot10^{-5}$ Ry. The spurious poles and zeros
	show a larger scattering. A zero of the true approximant at 
	$4.8371-i0.3177$ Ry, and a zero from the randomized data at $3.2222+i2.8388$ Ry not shown. Inset: Zoomed in around the imaginary axis.\\
	Bottom (b): Same as top, but here $\eta=10^{-3}$. Inset: Zoomed in around the imaginary axis.} 
	\label{g_addnoise}
\end{figure}

Note that for higher $N$ the number of defects also increases. As discussed in Sec. \ref{pzpairs}, the defects in the Pad\'{e} approximant should be sensitive to the multiplication of random numbers to the input data, while the physical poles should not. To test this effect, we multiply the input Green's function with random numbers as described in Sec. \ref{pzpairs}. The pole-zero distribution with and without multiplied random numbers are shown in the top panel of Figure \ref{g_addnoise} for the $N=16$ approximant using $\eta=10^{-6}$. The positions of the poles and zeros from the original data are listed in table \ref{gammaresidue}.
Except the (physical) pole ($\sim -0.6$ Ry below the Fermi level) the $14$ (unphysical) poles and zeros of the approximant are scattered in the complex plane. Before considering the multiplication with random numbers, four pole-zero pairs cluster around the imaginary axis (see inset Figure \ref{g_addnoise} top panel (a) - poles and zeros from the original data are black crosses/circles). The other poles and zeros are situated further out in the complex plane. As the original input data is multiplied by random numbers, the distribution of the poles and zeros changes. In particular, when $\eta=10^{-6}$ (top Figure \ref{g_addnoise}) the poles and zeros change their position illustrated with red pluses/triangles. All defects have changed their position while the true pole has only been shifted slightly compared with the defects. 
A stronger redistribution takes place when $\eta=10^{-3}$, as is seen in the bottom of Figure \ref{g_addnoise}. The true pole shows a larger shift than in the case of $\eta=10^{-6}$. 
Increasing $\eta$ to larger values than $\eta=10^{-3}$ gives an even larger shift of the true pole. 

Additional information about the approximant can be gained by investigating the residues of the poles, as explained in Sec. \ref{wall}. The residues calculated according to (\ref{wiqi}) can be seen in table
\ref{gammaresidue}, together with the position of their respective poles. Four of the residues $w_i$ ($i=1,2,4$ and $8$) have negative real parts and their respective poles can be ruled out as spurious. Residues with label $i=3$ and $6$ have have real parts that are several orders of magnitude smaller than the residue $i=5$, which has a 
real part close to 1. The residue of the $i=7$ pole has a large imaginary part compared with the other poles, indicating that the residue violates the restrictions given by the theorems cited in Sec. \ref{wall}.
 Hence we should conclude that the pole $q_5$ is our physical pole, and the rest are spurious. This conclusion is in perfect agreement with what was found above using random numbers on the input data. 

\begin{table}[h!]
\renewcommand{\arraystretch}{1.5}
\begin{tabular}{lllll}
$i$ & $Re(q_i)$ & $Im(q_i)$ & $Re(w_i)$ & $Im(w_i)$ \\
\hline
\hline
$\dag$ 1 & $1.1449469\cdot10^{-2}$ & $0.1360169$ & $-2.6324331\cdot10^{-13}$ & $-4.5109522\cdot10^{-13}$ \\
$\dag$ 2 & $-4.0764097\cdot10^{-2}$ & $0.3192908$ & $-1.3977842\cdot10^{-10}$ & $6.7675740\cdot10^{-11}$ \\
$\dag$ 3 & $-0.5137978$ & $1.111784$ & $1.0184415\cdot10^{-3}$ & $-1.3286583\cdot10^{-4}$ \\
$\dag$ 4 & $4.0205983\cdot10^{-3}$ & $0.2173124$ & $-1.2889500\cdot10^{-13}$ & $-2.2621590\cdot10^{-13}$ \\
$\star$ 5 & $-0.5958811$ & $-3.5719320\cdot10^{-6}$ & $0.9999138$ & $4.7717123\cdot10^{-5}$ \\
$\dag$ 6 & $-2.8055273\cdot10^{-2}$ & $0.1243891$ & $1.6093394\cdot10^{-12}$ & $-6.1991449\cdot10^{-13}$ \\
$\dag$ 7 & $1.547490$ & $2.5824822\cdot10^{-2}$ & $0.9696459$ & $0.1238033$ \\   
$\dag$ 8 & $2.441888$ & $-0.1386907$ & $-1.142889$ & $-0.1035025$ \\
\hline
\hline
\end{tabular}
\caption{Real and imaginary parts of the $N=16$ approximant poles with respective residues. The pole $i=5$ ($\star$) is a true
pole of the approximant, while the rest ($\dag$) are spurious. The poles can be seen plotted in Figure \ref{g_addnoise}.}
\label{gammaresidue}
\end{table}

\subsubsection{\textbf{k}-resolved Pad\'{e} approximation}

\begin{figure}[h!]
    \includegraphics[scale=0.5,clip=true]{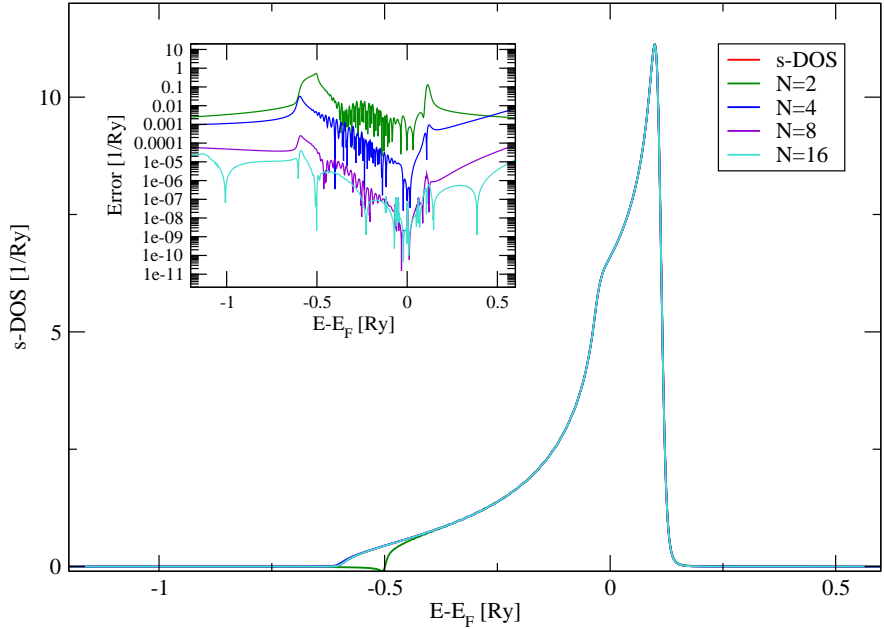}
	\caption{Density of states for solid hydrogen compared with approximants. The approximants were
	constructed using the $\ell m$-resolved \textbf{k}-dependent Green's function. Once the \textbf{k}-resolved Green's function was given for the horizontal density of states-contour the Bloch summation was performed. Inset: $Error$ as defined in Eq. (\ref{relerr}).}
	\label{hydkresdos}
\end{figure}

Repeating the same procedure as the one for the \textbf{k}-integrated Green's function but this time for the \textbf{k}-resolved $G_{RL,R'L'}(\textbf{k},i\omega_j)$, ($L=L'=\{\ell=0,m=0\}$, $R=R'$), using $N=2,4,8$ and $16$ number of input points produces the result shown in Figure \ref{hydkresdos}. As seen, the $N=2$ approximant fails to capture the DOS correctly at the bottom of the $s$-band. Higher order approximant provides a better fit.

\begin{figure}[h!]
    \includegraphics[scale=0.5,clip=true]{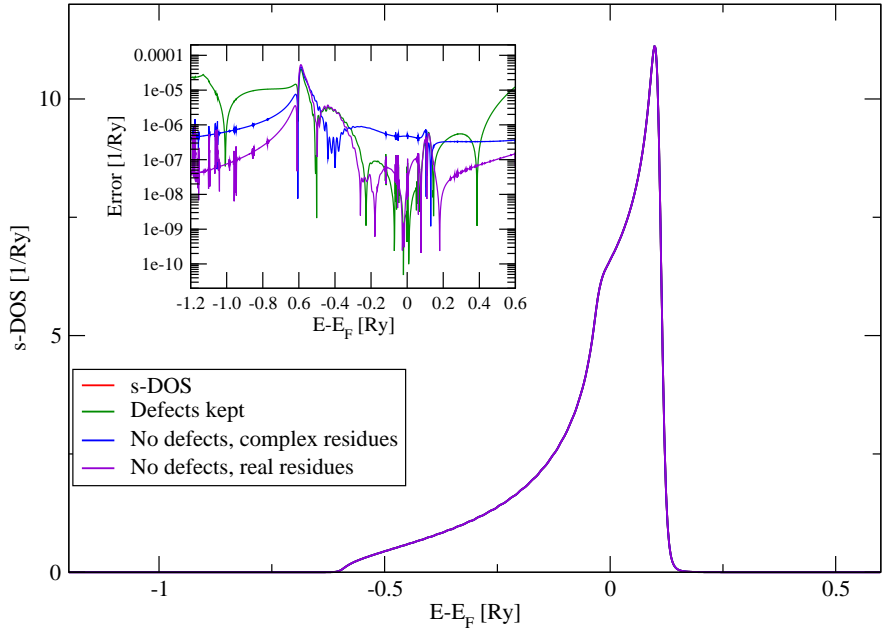}
    	\caption{Density of states and the $N=16$ approximant with defects (green) compared with approximants of the same
    	order, but reconstructed using
    	Eq. (\ref{wipi}). For one approximant (blue) the imaginary part of the residue was kept,
	for the other (violet) only the real part was used. Inset: $Error$ as defined in Eq. (\ref{relerr}).}
	\label{n16fullres}
\end{figure}

Using the information given by the inclusion of random numbers and from the residues, the approximant can be reconstructed using only the true poles, see Eq. (\ref{wipi}). We test this reconstruction on the $N=16$ approximant in Figure \ref{n16fullres}.
For each \textbf{k}-point the true pole was singled out. The reconstructed DOS is seen in Figure \ref{n16fullres} (blue line) plotted together with the $N=16$ approximant with the pole-zero pairs kept (green line) and the directly calculated DOS (red line). Both approximants are in agreement with the directly calculated DOS within the resolution of the figure. To get a better view of the discrepancies between the DOS, the error function defined in Eq. (\ref{relerr}) is plotted in the inset. As can be seen the reconstructed approximant (blue line) gives 
smaller errors for energies above $\sim 0.1$ Ry, and below $\sim -0.6$ Ry. In the case we enforce the approximant to have the analytic structure specified by the theorems cited in Sec. \ref{wall}, we demand the residues to be purely real, discarding the imaginary part. Reconstructing the approximant in this way gives rise to the violet line in Figure \ref{n16fullres}. It is noted that the error in this 
case is in general smaller than for the two former approximants, but not by a significantly large amount. Looking closer around the Fermi level it can be seen that the reconstructed approximants are not always giving smaller errors then the original approximant. However, the point of the above procedure is not to improve the general fitting, but rather to ensure that
no defects enter the approximant that would lead to non-analytic behavior.

\section{Summary and conclusion}
\label{conclusion}
Pad\'{e} approximants have long been in use for analytic continuation of spectral functions in condensed matter physics. This is since their rational polynomial form suits the analytical properties of the 
spectral functions of interest. The instabilities often encountered when using these methods should be attributed 
to the ill-posedness of analytic continuation in general, and not to the Pad\'{e} method itself. Indeed, as has been shown earlier \cite{be.go.00} the method can reach remarkable accuracy when the conditions concerning input precision and arithmetics are good enough. Typically this is not the case for electronic band structure calculations. However, our study shows that the analytic continuation can give acceptable results if special consideration is taken. The important issue to resolve is to be able to ensure that no spurious poles or zeros enter the complex half-plane in which the contour integrations and many-body equations are solved. As shown in this work, the above goal can be reached once the locations of the approximant poles and zeros are known, as one is then able to use the properties of the approximant to remove any unphysical features. 

\section{Acknowledgement}
A\"O and LV acknowledge the Swedish Research Council, the European Research Council, the Swedish Foundation for International Cooperation in Research and Higher Education (STINT), the Hungarian Scientific Research Fund (research project OTKA 84078) and the Anders Henrik Göransson Foundation for financial support.\\
In addition, A\"O acknowledges financial support offered by the Augsburg Center for Innovative Technologies and the generous hospitality of the chair of Theoretical Physics III, Center for Electronic Correlations and Magnetism, Institute of Physics, University of Augsburg.
The Swedish National Infrastructure for Computing (SNIC) and the MATTER Network, at the National Supercomputer Center (NSC), Link\"oping are acknowledged for computational resources.


\begin{thebibliography}{99}

\bibitem{ko.sa.06}
G. Kotliar, S. Y. Savrasov, K. Haule, V. S. Oudovenko, O. Parcollet and C. A. Marianetti, Rev. Mod. Phys. \textbf{78}, 865 (2006).

\bibitem{ka.ir.08}
M. I. Katsnelson, V. Y. Irkhin, L. Chioncel, A. I. Lichtenstein and R. A. de Groot, Rev. Mod. Phys. \textbf{80}, 315 (2008).

\bibitem{sa.ko.04}
S. Y. Savrasov and G. Kotliar, Phys. Rev. B \textbf{69}, 245101 (2004).

\bibitem{ch.vi.03}
L. Chioncel, L. Vitos, I. A. Abrikosov, J. Koll\'{a}r, M. I. Katsnelson and A. I. Lichtenstein, Phys. Rev. B \textbf{67}, 235106 (2003).

\bibitem{mi.ch.05}
J. Min\'{a}r, L. Chioncel, A. Perlov, H. Ebert, M. I. Katsnelson and A. I. Lichtenstein, Phys. Rev. B \textbf{72}, 045125 (2005).

\bibitem{an.sa.00}
O. K. Andersen and T. Saha-Dasgupta, Phys. Rev. B \textbf{62}, R16219 (2000).

\bibitem{an.je.94}
O. K. Andersen, O. Jepsen, and G. Krier, \emph{Lectures on Methods of Electronic Structure Calculation} (World Scientific, Singapore, 1994), p. 63.

\bibitem{vi.sk.00}
L. Vitos, H. L. Skriver, B. Johansson and J. Koll\'{a}r, Comp. Mat. Sci. \textbf{18}, 24 (2000).

\bibitem{vito.01}
L. Vitos, Phys. Rev. B \textbf{64}, 014107 (2001).

\bibitem{vito.07}
L. Vitos, \emph{The EMTO Method and Applications}, Computational Quantum Mechanics for Materials Engineers (Springer-Verlag, London, 2007).

\bibitem{an.ar.98}
O. K. Andersen, C. Arcangeli, R. W. Tank, T. Saha-Dasgupta, G. Krier, O. Jepsen, and I. Dasgupta, \emph{Tight-Binding Approach to Computational Materials Science}, edited by L. Colombo, A. Gonis, and P. Turchi, MRS Symposia Proceedings No. 91 (Materials Research Society, Pittsburgh, PA, 1998), p. 3–34.

\bibitem{zw.an.09}
M. Zwierzycki and O. K. Andersen, Acta Phys. Pol. A \textbf{115}, 64 (2009).


\bibitem{vi.se.77}
H. J. Vidberg and J. W. Serene, J. Low. Temp. Phys. \textbf{29}, 179 (1977).

\bibitem{be.go.00}
K. S. D. Beach, R. J. Gooding and F. Marsiglio, Phys. Rev. B \textbf{61}, 5147 (2000).

\bibitem{baker.75}
G. A. Baker, \emph{Essentials of Pad\'{e} approximants} (Academic Press, New York, 1975).

\bibitem{so.ak.11}
D. Sokolovski, E. Akhmatskaya. S. K. Sen, Comp. Phys. Comm. \textbf{182}, 448 (2011).

\bibitem{wall.48}
H. S. Wall, \emph{Analytic theory of continued fractions} (Chelsea, New York, 1948).

\bibitem{wa.we.44}
H. S. Wall and M. Wetzel, Trans. Am. Math. Soc. \textbf{55}, 373 (1944).

\bibitem{pe.ca.95}
M. S. Petkovic, C. Carstensen, M. Trajkovic, Numer. Math. \textbf{69}, 353 (1995).

\bibitem{ki.vi.07}
A. E. Kissavos, L. Vitos and I. A. Abrikosov, Phys. Rev. B \textbf{75}, 115117 (2007).

\bibitem{pe.wa.92}
J. P. Perdew and Y. Wang, Phys. Rev. B \textbf{45}, 13244 (1992).

\end{thebibliography}
\end{document}